# Influence of surfactants on the electrical resistivity and thermopower of Ni nanoparticles

**Netram Kaurav,[1] Gunadhor S. Okram[2*] and V. Ganesan[2]**

[1]Department of Physics, Government Holkar Science College, A. B. Road, Indore 452001, MP (India)

[2]UGC-DAE Consortium for Scientific Research, Khandwa Road, Indore 452001, MP (India)

*Corresponding author. E-mail: okram@csr.res.in, okramgs@gmail.com

Fax +91-731-2462294

**Abstract**

Compacted pellets of nanocrystalline nickel (NC-Ni) of average particle size ranging from 18 to 33 nm were prepared using a variety of surfactants. They were characterized well and were studied on the influence of the surfactants on the electrical resistivity and thermopower in the temperature range 5 to 300 K. It was found that the type of the surfactant used dominates over the average particle size in their electrical transport and the detail transport behaviors have been discussed. Moreover, the observed thermopower and resistivity features were contrasting compared to what are normally seen in the well-known materials. They are interpreted as indicative of attractive features these surfactants for the design of nanostructured thermoelectric materials with enhanced thermoelectric figure of merits.

# I. INTRODUCTION

Nanocrystalline (NC) metallic nanoparticles such as the ferromagnetic metals Fe, Co and Ni have been widely studied because of their use in many applications including magnetic fluids [1], magnetic recording media [2], biomedical applications [3, 4] and catalysis [5]. In particular, nanostructured Ni particles have important applications ranging from magnetic sensors [6] and memory devices to biomolecular separations [1, 3]. Synthetic protocols are believed to be important in order to take maximum advantage of those novel properties which are often affected by particle size, shape and crystalline phase. So far, numerous physical and chemical methods have been used to produce metal nanoparticles, such as metal evaporation-condensation [7], laser ablation technique [8], electrochemical methods [9], sonochemical synthesis [10], metal salt reduction [11] and neutral organometallic precursor decomposition [12]. Generally, chemical synthesis methods have the advantages of simplicity and low cost compared with physical approaches. Moreover, it allows production of large quantities of nanoparticles of defined forms with a fairly narrow particle size distribution.

An important feature in the production of the nanoparticles is the ability to keep them physically isolated one from another preventing reversible aggregation. The stability of the nanoparticles is commonly achieved using different protective molecules, which bind on the surface of nanoparticles avoiding their aggregation [13, 14]. Interestingly, the polyol process uses a poly-alcohol (ethylene glycol-EG, for example) as both solvent and reducing agent to produce nanoparticles from metallic cationic precursor [15-17]. In this process, the polyol itself can act as protective agent to avoid particles agglomeration and growth.

Furthermore, one of the important factors of the nanoscale regime is the presence of a large percentage of surface atoms and/or a large amount of grain boundaries (GBs) that give rise to the unusual properties of NC-materials compared to conventional polycrystals or single crystals with the same chemical composition [18]. It has been established adequately that the electrical resistivity ($\rho$) is not determined by the material alone but by its size as well [19-24]. Fascinating aspects of nanostructures in this regard are spatial confinement of carriers and corresponding change in carrier density of states [18, 25, 26]. With nanoparticles as quantum wells, phonon dispersion and group velocities change due to spatial confinement induced by GBs [25]. Phonon confinement affects all phonon relaxation rates and should affect thermoelectric transport as in electrical transport, wherein GB disorder prevails [18, 26, 27]. The scattering due to small size and more GBs show up as the residual resistivity at $T \leq 20$ K in nearly all metallic solids. If the NC-material is comparatively more disordered [28] or has proximity effect due to its substrate or matrix [29], the resistivity can show localization instead of causing a constant residual value. Nearly comparable situation of matrix-dependence could be applicable in the solution-method of preparation of samples and in-depth investigations are essential when one uses different surfactants for generating the nanoparticles in their compacted forms. In this situation, when a surfactant molecule or its head sticks to the nascent metal surface in restricted small size, energy levels or Fermi energy of the electrons in the equilibrated state could be modified. This equilibrated Fermi energy will be decided by the final state of electronic distribution in contact with a particular surfactant since they should have distinct interactions [18, 24].

In order to investigate some of these aspects, we have chosen ethylene glycol (EG), $CH_2OHCH_2OH$, diethylene glycol (DEG), $OHCH_2CH_2)_2O$, polyethylene glycol (PEG), $H-(O-CH_2-CH_2)_n-OH$, oleic acid (OA), $CH_3(CH_2)_7CH=CH(CH_2)_7COOH$, urea (UA), $CO(CH_2)_2$ and tetrapropylammonium hydroxide (TPAH), $(C_4H_9)_4NOH$ as surfactants or stabilizers for preparing NC-Ni. From the chemical point of view, EG with both polar and non-polar ends is miscible with both water and organic compounds. DG also has similar properties as EG but it has an oxygen double bond. PEG may have chains emanating from a central core group. OA is long chain organic compound having a double bond, which is mostly centre for reaction with other compounds. UA with a carbonyl group is soluble in water but is neither acidic nor alkaline. TPAH is soluble in organic solvents. How will they influence the meta-stably small but metallic nanoparticles, how they create barriers among the latter and hence influence the electronic transport in this composite-like nanostructures? are a matter of utmost interest for use in future interconnects and materials properties.

The aim of this work is therefore to explore the influence of these surfactants on the magnetoresistive and thermopower behaviors at low temperatures of nickel nanoparticles by using a synthetic protocol that uses EG and DEG as both solvents and reducing agents, wherein a nickel acetylacetonate is thermally decomposed to generate the nanoparticles. Preliminary results were reported earlier [30]. We present here the detail results and analysis. Employing EG and/or DEG not only simplify the preparation processes but also provide interesting routes to prepare high-quality nanoparticles of controlled structure, size and morphology. The present findings in overall suggest the significant influence of surfactants on contrasting behavior of resistivity and thermopower, thereby, promising approach towards the realization and/or design of thermoelectric materials in enhancing the thermoelectric figure of merit.

## II. EXPERIMENT

### A. Sample preparations

Nanocrystalline Ni (NC-Ni) samples were prepared by refluxing the solution of nickel acetate, NiAc, $Ni(Ac)_2.4H_2O$ in ethylene glycol ($CH_2OH-CH_2OH$)/ diethylene glycol ($C_4H_{10}O_3$). A series of NC-Ni samples were prepared in different capping agents. Typically, 0.1 M (1.2424g) of NiAc was dispersed in 50ml of EG/DEG and refluxed for about one hour at ~190$^O$C. As the reflux was in progress, after about half an hour, initially light green color solution turned milky greenish color and slowly turned dark brown color wherein black Ni particles were dispersed. The surfactant(s) was/ were added after two hours of stirring of the polyol mixed with NiAc. After this, two hours more stirring was done. Then, resultant solution was refluxed for two hours.

A few milligrams of OA as surfactant were used for Ni-1 sample. A few milligrams of urea and PEG were added as surfactants for Ni-2 sample. Similarly, OA and tetrapropylammonium hydroxide (TPAH) were used as surfactants for Ni-3 sample. Ni-4, Ni-5 and Ni-6 samples were prepared in diethylene glycol only (Ni-5) instead of ethylene glycol using TPAH alone (Ni-4) and OA alone (Ni-6) as surfactants. It was noted that if we add more quantity of oleic acid the formation of oleic acetates or corresponding acetates was there.

### B. Characterization and other physical measurements

The characterization of particle size, structural and crystallographic nature of the nanoparticles forms an essential part in the analysis of the data. The NC-Ni samples used in this investigation are polycrystalline in nature. This including the determination of the average particle sizes has been established from x-ray diffraction (XRD) data. The nanostructure was also investigated using atomic force microscopy (AFM). Magneto-resistivity measurements were performed by standard four probe technique using a commercial cryostat (OXFORD Instruments, Inc., UK) in the temperature range 5 to 300 K and up to 8 Tesla magnetic fields. Thermopower measurements down to liquid helium temperature (5 – 300 K) were carried out using our homemade setups [31]. Correction for the reference copper in thermopower measurements was done, which is crucial for the accurate determination of the measured thermopower.

# III. CHARACTERIZATION
## A. X-ray diffraction

Figure 1 shows the XRD patterns of the samples with different surfactants. Diffraction pattern exhibited four distinctive peaks near 44.58°, 51.74°, 76.4° and 93° which can be indexed as (111), (200), (220) and (311) reflections, respectively of fcc nickel (JCPDS#). It is clear that all the peaks correspond to fcc structure of the Ni metal and there is no sign of formation of hcp Ni. Upon exposure to air NC-Ni synthesized via thermal decomposition can potentially oxidize to NiO [32, 33]. No nickel oxide peaks were observed in the nickel nanoparticles synthesized in this work owing to the surrounding capping layer on the surface of the nanoparticles. The extra peak (as indicated by stars (*) in figure 1) seen in some samples is assigned to the presence of hydroxide (OH) which were removed by keeping the sample in an oven at a temperature between 60°C to 80°C for two days before electrical transport measurements. The calculated crystallite size ($D$) using Scherrer equation in XRD patterns is found to be ~18 to 33 nm, showing the influence of surfactants on size. Here, the $D$ was calculated from full width at half maximum (FWHM) of most intense diffraction peak (111).

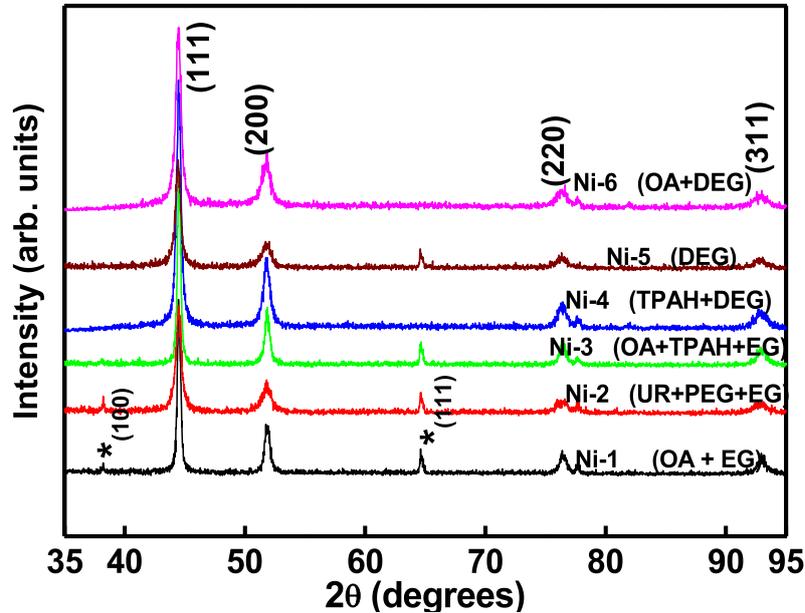

**Figure 1.** Shows the XRD patterns of the samples with different surfactants. The extra peak (as indicated by *) seen in some samples is the presence of hydroxide (OH).

## B. Atomic Force Microscopy

To understand the particle size and microstructure better, we have taken the AFM image of samples Ni-2, Ni-5 and Ni-6 of different surfactants. For this, their representative AFM images are presented in Fig. 2(a-c, left panel). The size $D$ estimated from the images of 50-100 nanoparticles by noting down the frequency of particles is found to be 34±1, 42±1 and 44±2 nm nm for Ni-2, Ni-5 and Ni-6, respectively. For this, a bar chart was plotted and a Gaussian fitting was performed (Figure 2(a-c), right panel). It is seen that the grain size obtained from AFM are different as estimated from XRD for the same sample (table 1), which is related to the sensitivity of each technique. In XRD, even though the area involved for observation is global, it provides average coherence length related to local periodicity inside the grains. The average coherence length is thus usually smaller than actual physical size seen through the AFM. This gives a real space distribution of grains in a small region. However, here we have used XRD estimated particle size to interpret our results.

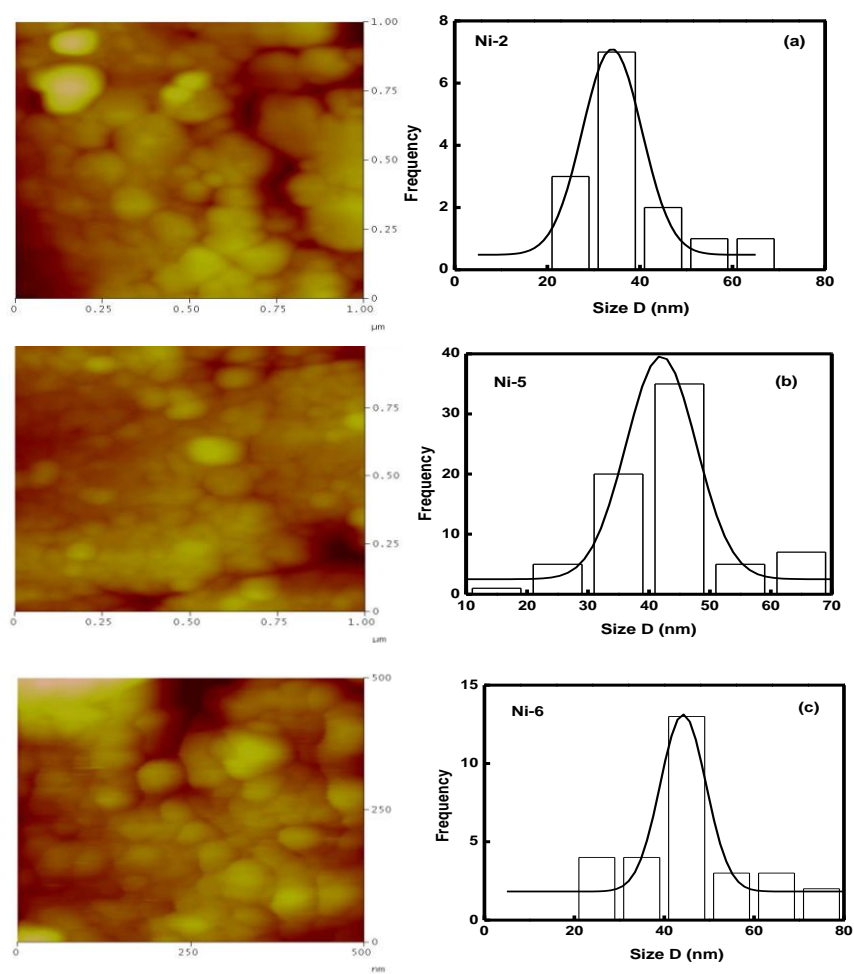

**Figure 2.** AFM images (left panel) and particle size distribution (right panel) of (a) Ni-2 [PEG + UR], (b) Ni-5 [DEG], (c) Ni-6 [DEG+OA] samples.

**Table 1** The details of overall chemical preparation of Ni nanoparticles and particle size (D) from XRD.

| Sample | Solvent | Capping Agent (g) | D(nm) |
|---|---|---|---|
| Ni-1 | EG | OA (0.05) | 31 |
| Ni-2 | EG | PEG (2g) + Urea-(1g) | 18 |
| Ni-3 | EG | OA(0.05)+TPAH(0.05) | 32 |
| Ni-4 | DEG | TPAH (0.05) | 23 |
| Ni-5 | DEG | -- | 21 |
| Ni-6 | DEG | OA (0.05) | 20 |

### III. RESULTS AND DISCUSSION
#### A. Electrical resistivity

Figure 3 shows the temperature dependence of the electrical resistivity $\rho_n(T)$ of the compacted NC-Ni samples with $D \sim$ 18-32 nm in the temperature range from 5 to 300 K. The prepared powder was compacted into pellets by applying a pressure of about 2GPa. This gives an average sample density of about 85% of the bulk density that agrees well with those reported earlier [24]. The electrical resistivity (Fig. 3) exhibits the strong influence of different surfactants. The $\rho$ values at 300 K and 5 K are seen to have systematic change (Fig.3, inset). Interestingly, the samples prepared with EG as the base solvent-cum-surfactant show much higher values of resistivity compared to those of DEG indicating that the electron localization due to EG at the grain boundaries and EG itself causes significant reduction in the conductivity compared to its bulk counterpart as well as that caused due to DEG. This would be correlated with the presence of DEG with oxygen double bond that probably makes a situation that electrons are more delocalized compared to that in EG presence. For the sample with OA in EG (Ni-1, 31nm) or PEG and urea together in EG (Ni-2, 18nm), $\rho$ is high and nearly the same. This indicates that influence of OA with a double bond and a long carbon chain provides least delocalization among all other samples irrespective of its larger particle size compared to others especially the Ni-2 sample. Comparatively, urea with carbonyl group and PEG with long chain, in addition its (Ni-2) relatively much smaller particle size (18nm) compared to 31nm (Ni-1) do not lead to much higher resistivity as in Ref. [27]. This would indicate that the electron localization with OA is much stronger than that of combined effect of urea and PEG. In contrast, TPAH enhances the electron delocalization that in turn nullified the role of OA in 32nm (Ni-3) sample. This leads to a reduction in $\rho$ by an order.

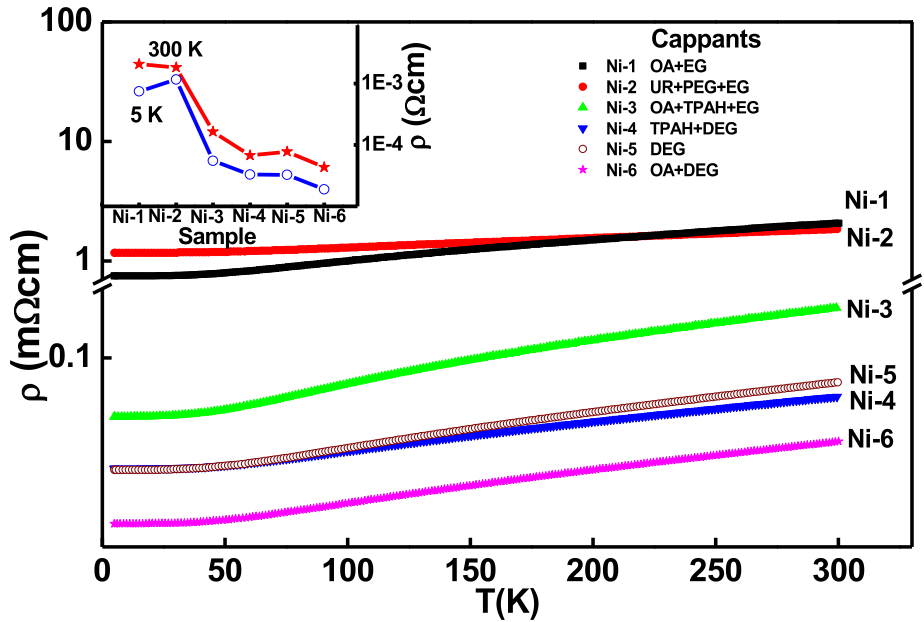

**Figure 3** Variation of electrical resistivity with temperature of compacted NC-Ni samples. Inset shows variation of ρ(300K) and ρ(5 K) with different surfactants.

However, the situation changes completely when DEG is used as solvent-cum-surfactant wherein $\rho$ is drastically smaller than that of others that defies their (Ni-4, Ni-5, Ni-6) relatively much smaller size (23, 21, 20nm, respectively) in disagreement of enhanced $\rho$ with size reduction [27]. This is in addition to the modified and enhanced roles of *either OA or TPAH* in the delocalization of the electrons in the presence of DEG in contrast to that of ethylene glycol. Clearly, the grain boundary region with DEG is relatively very favorable with better electron delocalization compared to those in ethylene glycol, as the resistivity features and values suggest. Comparing further the surfactants OA and TPAH, we observed that OA degrades the delocalization i.e. $\rho_{Ni5}>\rho_{Ni4}$ at least at 300K, but TPAH enhances the electron delocalization due to DEG. This leads Ni-6 sample to be the one with the smallest resistivity despite its smallest particle size sample, and is in complete contradistinction from EG (Ni-1, Ni-2, Ni-3) samples, wherein converse is true.

Even with these significant change in ρ, the slopes remain positive throughout the temperature range in spite of their high values. The large ρ nature of the compacted Ni-NP pellets is attributed to the disorder in grain boundary regions that effectively represent a series-resistor network [24, 27] and their interactions with the surfactants. The positive slopes however rule out any drastic disorder in the system that can give rise to effects such as electron localization [29, 34]. The relatively more resistive nature of compacted NC-Ni pellets with EG compared to that of DEG is attributed to the disorder in the grain boundary regions combined with the better more localizing nature of the former. This consequently leads to enhanced scattering even at higher particle size *D*. This trend clearly indicates the significant role of grain boundaries and surfactants affecting the smooth flow of charge carriers. The other possibility comes from confinement of phonons, which is also enhanced due to the grain boundaries (GBs) present in the system [18, 27]. Previously, it has been shown that the phonon confinement affects all phonon relaxation rates and should affect thermal transport, wherein GB disorder prevails [18]. With these entire features, two other important points to be noted are that resistivity of samples Ni-1 and Ni-2 crosses near 220K while those of samples Ni-4 and Ni-5 crosses near 20K. They indicate that the slopes also are affected by urea and tetrapropylammonium hydroxide as surfactants. The present NC-Ni samples thus provide an opportunity to understand these processes based on the phonon confinement, which might be one of the important factors responsible for the observed anomalous nature of thermopower presented in the later part of this paper.

**B. Magnetoresistance**

The magnetoresistance (MR) properties of the synthesized NC-Ni samples were also investigated. The magnitude of MR is defined as: $[\Delta R/R(0)] = [R(H)-R(0)]/R(0)$, where $R(H)$ and $R(0)$ are the resistances at a given temperature in presence and absence of a magnetic field, H respectively. Figure 4 shows the temperature variation of %MR in NC-Ni samples in the presence of applied magnetic field of 8 T. It has been found that the %MR for samples with EG (Ni-1, Ni-2 and Ni-3) as one of the surfactants exhibit quite dramatic response with temperature. MR of Ni-1 and Ni-3 samples decreases with increasing temperature and becomes negative at higher temperatures (150K and 240K, respectively). Ni-1 sample especially show the largest range of MR values for temperature range studied. When TPAH surfactant is introduced (Ni-3), MR evolves into higher values at lower temperatures turning negative marginally above 240K. However, the use of UR and PEG with EG (Ni-2), in overall smaller %MR was found and negative values throughout the temperature range we studied. Thus, all the samples made with EG shows negative value that is significantly large near room temperature. Conversely, samples made in DEG only or with TPAH and OA separately exhibit very marginally increasing trends with positive MR values. This overall scenario corroborating the significantly anomalous electrical transport behavior seen in resistivity clearly reflects the usually observed negative and positive MR in the interfaces [35] or nanocomposite fibers [36], especially the organic semiconductor devices [37].

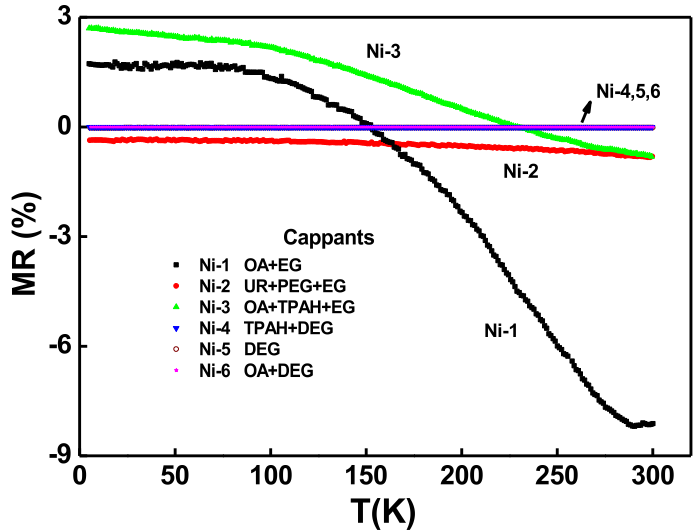

**Figure 4** Temperature dependence of percentage of magneto resistance (%MR) for an applied field of 8 T.

The bigger particle size for EG samples wherein the presence of a sufficient concentration of ferromagnetic (FM) clusters of Ni is expected seems to suffice for the observation of MR in these samples. Due to the randomly distributed moments of FM clusters as well as the large grain contact resistance, magnetic disorder may play a key role in electron localization. Under an applied magnetic field, the orientation of random FM clusters are forced to align uniformly so that the magnetic disorder is reduced, which favors the electron delocalization and results in a significant drop of resistivity. On the contrary, negligible MR has been observed for the NC-Ni prepared in DEG (Ni-4 to Ni-6) as solvent-cum-surfactant. The insensitivity of these samples for magnetic field is also consistent in the picture as augmented above. These samples with the smaller particle size as compared to that of EG samples probably have reduced the ferromagnetic clusters of Ni without effective magnetism. Hence, such an observation might suggest superparamagnetic state in these samples [38].

## C. Thermopower

Figure 5 displays the temperature dependence of the thermopower $S_n(T)$ of the NC-Ni samples in the temperature range from 5 to 300 K. The $S(T)$ data of Ni bulk (99.99%) is also included for comparison, which shows negative sign, a broad phonon drag peak around 50K and its values lies within -20 $\mu VK^{-1}$. These features have also been exhibited by the NC-Ni samples but sign of thermopower is found to be positive at low temperatures and magnitude of S changes with the type of surfactant. As the temperature increases, all samples display positive sign up to ~35 K (except for Ni-2 sample) turning negative above this temperature. On close observations, it is seen that these features evolve into a very dominant role of surfactants in defiance to size-dependent S observed earlier [18]. EG (Ni-2) sample prepared in polyethylene glycol with urea exhibits smallest S and broader positive S region at low temperatures but showed the second maximal resistivity at 300K (Figure 2) in spite of the fact that its particle size (18nm) is the smallest compared to those of Ni-1 (31nm) and Ni-3 (32nm) among the samples prepared in ethylene glycol, and three others prepared in diethylene glycol viz, Ni-4 (23nm), Ni-5 (21nm) and Ni-6 (20nm). Especially, for the latter three samples, as per the relatively quite lower resistivity values, their *S* should have been smaller than that of Ni-2. This however is on the contrary to the common perception of the higher the resistivity the bigger the thermopower [39]. In overall, the general trend of these nanoparticle samples seen in resistivity is not followed in thermopower. Regarding the hump, it turns out to be more prominent and sharper for Ni-3 (32nm), but quite flattened for Ni-2 (18nm). These contrasting features exhibited in the thermopower and resistivity that is contradictory to what is normally seen in metals and alloys [39] and even in semiconductors [40] might indicate some attractive features related to designing thermoelectric materials with enhanced the thermoelectric figure of merit [41]. They therefore suggest the critical roles of ethylene glycol, diethylene glycol, polyethylene glycol, oleic acid, urea and tetrapropylammonium hydroxide as surfactants and their interactions with the nanoscale metallic surfaces.

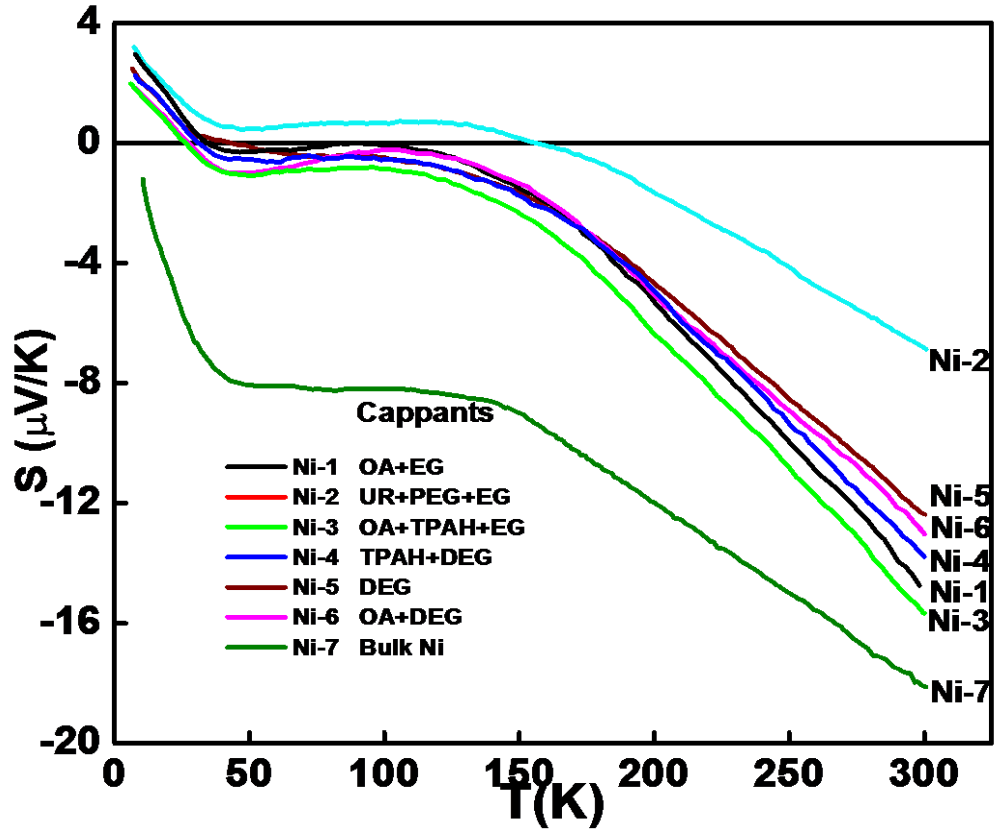

**Figure 5** Variation of thermopower of compacted NC-Ni samples with temperature. The thermopower of nickel (Bulk) is also shown.

These features might indicate a substantial change in the band structure or the conduction mechanism at the nanoscale. The flipping of sign of the thermopower in these NC-Ni samples indicates, in principle, involvement of both the electrons and holes in the electrical transport. From ~150 to 300 K, magnitude of S, |S| in general increases systematically as the temperature rises, indicating that the diffusion thermoelectric transport prevails in the high temperature regime suggestively with electrons as majority charge carriers. Our results clearly indicate distinct behavior of the NC-Ni samples at low temperatures, associated with sharp onset of a negative phonon drag peak. This is intriguing as it is distinct from that of the bulk (electron-like throughout the temperature range), since the thermopower in general is dependent on the topology of the Fermi surface and consequently closely related with the energy (E) and phonon wave vector (*k*) dependence of the relaxation time, $\tau(E,k)$.

In the low temperature limit, the carrier relaxation time is limited by impurity scattering. From the extrapolation of the electrical resistivity of Ni-nanoparticles (Figure 3) to zero temperature, it turns out that the present samples correspond to distinct value of resistivity at zero temperature. This means that there is an essential presence of *equivalent* impurity scattering in the present samples. Thus for low temperatures, where phonons have not yet started to play a significant role, the electrical resistivity is proportional to inverse of relaxation time. For a three-dimensional system, the electron diffusion contribution to the thermopower is distinguished by its simple linear temperature dependence. The low as well as high temperature phonon drag and diffusion mechanisms have been adequately discussed for metals [39, 42]. In the low temperature regime, the thermopower follows the relation

$$S = AT + BT^3. \tag{1}$$

Here the first term in the isotropic relaxation time approximation represents the contribution from diffusion component ($S^d$) and is linear in *T*. The second term corresponds to phonon drag contribution ($S^{ph}$) at low temperature, analogous to the specific heat that follows Debye $T^3$ law. On the other hand, at higher temperatures, conventional diffusion and phonon drag component to *S* is written as

$$S = aT + b/T. \qquad (2)$$

Here, *A* (or *a*) and *B* (or *b*) are diffusion coefficient and phonon drag coefficient at low (high) temperatures, respectively. In order to elucidate these mechanisms and to extract the diffusion coefficient (*a* or *A*) and phonon drag coefficient (b or *B*) from the observed thermopower of NC-Ni samples, we have plotted *S/T* versus $T^2$ and *S/T* versus $T^{-2}$ for the low and high temperature regions in Figures 6 and 7, respectively. We have plotted in turn the resultant A & B (Figure 6, inset) and a & b (Figure 7, inset) coefficients versus surfactants. It

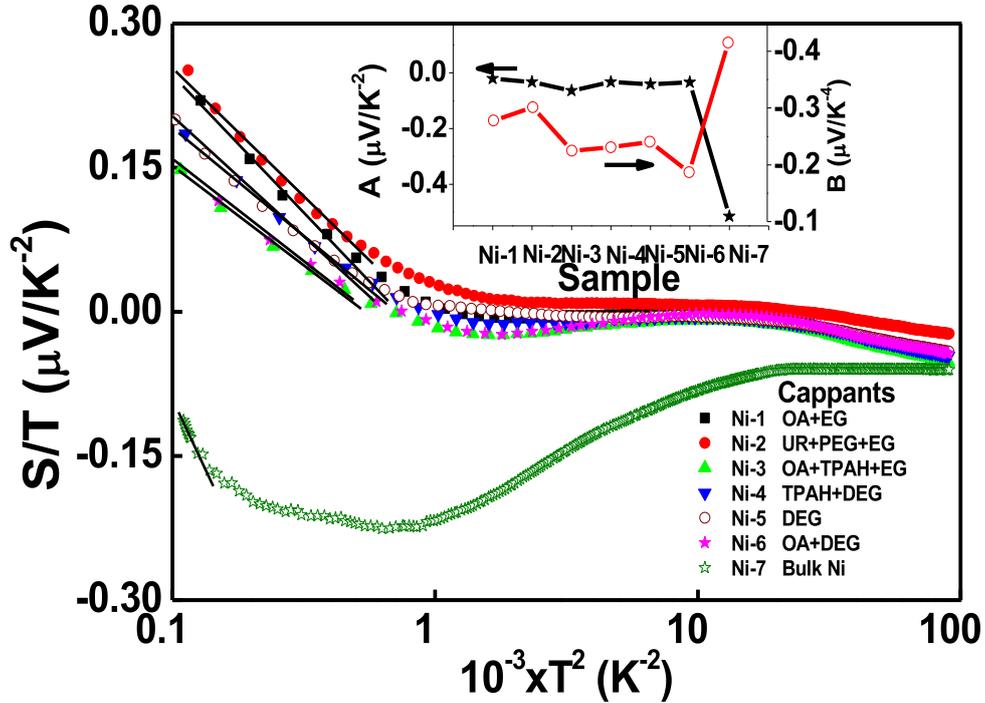

**Figure 6** Diffusion and phonon drag contributions of various NC-Ni samples. *S/T* is found to be linear with $T^2$ below 50 K. The inset shows the coefficients *A* and *B* in Eq. (1) plots against surfactants.

is seen that the magnitude of coefficient *A* in the low temperature regime is higher as compared to bulk Ni sample. Conversely, |*B*| is found to be lower as compare to bulk Ni sample. That is, these (diffusion and phonon drag contribution) coefficients operate

oppositely to each other as expected. For the high temperature regime, the value of coefficient *a* is found to higher for sample Ni-2 (*D* = 18 nm) as compared to the other samples. In other words, the phonon contributions or confinement is prominent in this sample compared to other bigger nanoparticle samples.

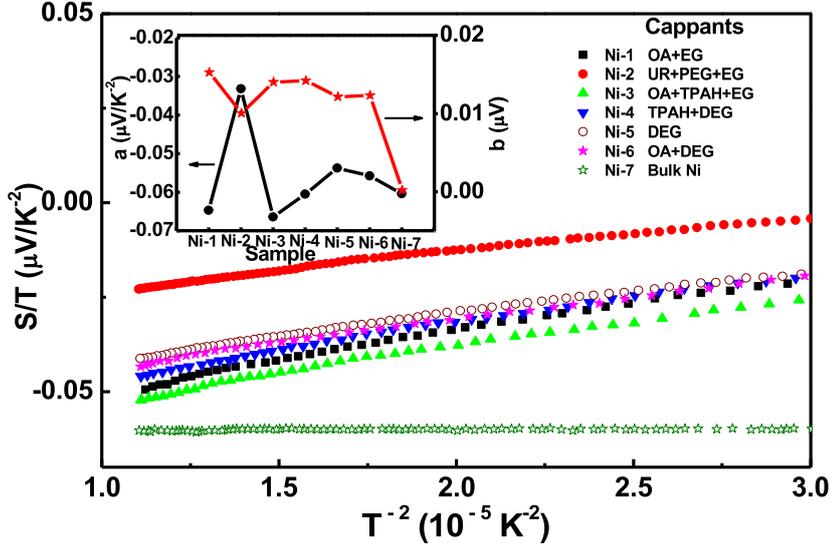

**Figure 7** Diffusion and phonon drag contributions of various NC-Ni samples. *S*/*T* is found to be linear with $1/T^2$ above 150 K. The inset shows the coefficients *a* and *b* in Eq. (2) plots against surfactants.

To further elucidate the speculation on the phonon confinement effect in the present NC-Ni samples, we recall that the effect of electron-phonon confinement would manifest itself as a broad hump in *S* at low temperatures for highly disordered GBs scattering [18, 43]. In the present case the low temperature limited resistivity for all measured NC-Ni samples is less than 2 mΩ cm (Fig. 3), suggesting that the phonon confinement effect in thermopower is expected to be noticeable in these compacted nanoparticles. Significantly disordered GBs in nanocrystalline metals introduce numerous interfaces leading to extra scattering centers due to the size effect for electrons and phonons [18, 43]. To get extra contribution of phonon drag and diffusion thermopower due to disorder at grain boundaries and particle size, difference in *S*, $\Delta S = S_n - S_{bulk}$ is plotted in Figure 8. It is seen that the absolute value of Δ*S* is significantly larger for all the samples compared to *S* i.e. there is significant enhancement in *S* in NC-Ni samples with a systematic variation of Δ*S* as a function of temperature. The higher Δ*S* is found to be for Ni-2 (*D* = 18 nm) sample as expected. We interpret this to the random nature of the GBs and surface atoms. As the GBs and surface atoms increase with decrease in particle size, |Δ*S*| systematically increases and consequently the diffusion and phonon-drag contributions become relatively more prominent than those in bulk |S|. Notably, Δ*S* exhibit broader peaks with a marginal shift in their positions to lower *T* as *size* decreases. These results clearly indicate dominant role of GB disorder with additional factors due to the different surfactants, which influence scattering of phonons and electrons in a significant way.

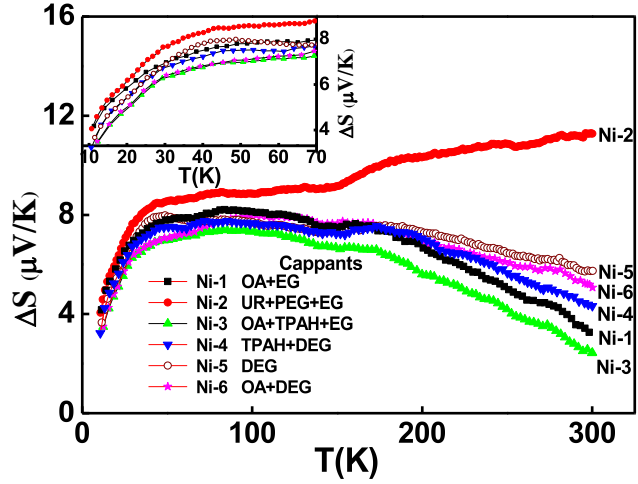

**Figure 8.** Difference thermopower, Δ*S*, of compacted NC-Ni samples.

### IV. CONCLUSIONS

In conclusion, several compacted Ni nanoparticles of average particle size (*D*) ranging from ~18 to 33 nm have been prepared using ethylene glycol, diethylene glycol, polyethylene glycol, oleic acid, urea and tetrapropylammonium hydroxide as surfactants. They were characterized using XRD, AFM, electrical resistivity, magnetoresisitance and thermopower to elucidate the critical role of surfactants. The observed electrical resistivity $\rho_n(T)$ data for all samples are typical of a good metal and have a fairly linear temperature dependence of resistivity down to about 50 K. It is found that the samples prepared with ethylene glycol as the base solvent-cum-surfactant show much higher values of resistivity compared to those of diethylene glycol indicating that the *grain-contacts are relatively much conductive in the latter samples than those of former ones* and eventually produces negative and significant values of magnetoresistance. In contrast, the trends in thermopower are different from those seen in resistivity that is contradictory to what is normally seen in most of the known materials. This might suggest quite attractive features related to design aspects of thermoelectric materials that may help in enhancing the thermoelectric figure of merit and hence important roles of surfactants.

## ACKNOWLEDGEMENTS

Authors gratefully acknowledge the assistance rendered by Joffy of Barkatullah University, Bhopal, M.P. in the sample preparation, M. Gupta and R. Rawat, UGC-DAE CSR, Indore for XRD, resistivity and magneto resistance data collection, respectively.